# Thermal Transport in SiC with Intrinsic Defects and Mg Transmutation Products

Chen Shen,[1,*] Yang Su,[2,*] Maciej P. Polak,[1,*†] Rafi Ullah,[3] Nuohao Liu,[1] Mary Alice Cusentino,[4] Dane Morgan,[1] and Izabela Szlufarska[1,‡]

[1]*Department of Materials Science and Engineering, University of Wisconsin-Madison, Madison, Wisconsin, USA*
[2]*College of Physics, Jilin University, Changchun, Jilin, China*
[3]*Institute of Physics, Hiram College, Hiram, Ohio, USA*
[4]*Sandia National Laboratories, Albuquerque, New Mexico, USA*

Silicon carbide is a leading candidate material for advanced nuclear energy systems, but irradiation-induced defects and transmutation products can severely degrade its thermal conductivity. In fusion environments, Mg is predicted to be a major solid transmutant in SiC, yet it is not well understood how different Mg-related defects affect phonon transport. Here, we develop a machine-learning interatomic potential, MLIP4SiC-Mg, for 3C-SiC containing intrinsic point defects, Mg-related defects, and Mg-defect complexes. The potential is trained on a large DFT dataset and reproduces DFT energies, forces, equation-of-state behavior, phonon dispersions, and lattice thermal conductivities with near-DFT accuracy. Combined with Green-Kubo molecular dynamics, force-error correction, and a resistance-based treatment for dilute defective systems, MLIP4SiC-Mg enables quantitative thermal-conductivity calculations in large defective supercells. The corrected thermal conductivity of pristine 3C-SiC is 421 W $m^{-1}$ $K^{-1}$ at 300 K, in good agreement with available experimental data. All defects considered strongly reduce thermal conductivity, but their scattering strengths are highly configuration dependent. $V_C$ and $Mg_{TC}$ act as strong phonon scatterers, whereas isolated $Mg_{Si}$ is comparatively weak. Residual thermal resistivity analysis shows that defect-induced thermal resistance is not strictly linear with concentration and should be treated as an effective temperature- and concentration-dependent scattering metric. $Mg_{Si}$-$V_C$ clustering enhances scattering relative to isolated $Mg_{Si}$, but reduces the total excess resistance relative to spatially separated $Mg_{Si}$ and $V_C$ defects. These results clarify the configuration-dependent role of Mg transmutation in irradiation-degraded SiC and provide an atomistic framework for quantifying defect-controlled heat transport in nuclear ceramics.

## I. INTRODUCTION

Silicon carbide (SiC) is widely regarded as a promising material for harsh nuclear environments because of its excellent thermal, mechanical, and chemical stability. Its high melting point, high thermal conductivity, low neutron absorption cross section, and strong resistance to oxidation and corrosion make it attractive for structural applications under high temperatures and intense irradiation [1]. In fusion systems, SiC-based materials have been considered for first-wall and blanket components because of their dimensional stability and low activation [2]. In fission systems, SiC is a key constituent of tristructural isotropic (TRISO) fuel particles for high-temperature gas-cooled reactors, where it provides mechanical integrity and serves as an effective barrier to fission products [3, 4]. Beyond TRISO fuels, SiC has also been proposed for light-water reactors, either in fully ceramic microencapsulated (FCM) fuel designs [5] or as an accident-tolerant alternative to conventional zirconium alloys in $SiC_f$/SiC composite cladding [6].

Thermal conductivity is one of the most important properties governing the reliability of SiC-based components, because heat removal directly affects temperature gradients, thermal stresses, and service lifetime. In nuclear systems, the thermal conductivity of cladding is closely tied to operational safety. High-conductivity cladding helps maintain a more uniform temperature profile during normal operation and reduces the temperature drop between the fuel and the cladding. Under off-normal conditions, such as a loss-of-coolant accident, efficient heat transport can also delay temperature rise after shutdown and improve accident tolerance [6, 7].

In nuclear applications involving significant neutron exposure, SiC can experience irradiation damage in both fission and fusion systems, including the SiC layer of TRISO fuel particles, $SiC_f$/SiC fuel cladding, and core structures in fission reactors, as well as first-wall and blanket components in fusion reactors [1, 8, 9]. In 3C-SiC, neutron irradiation can affect thermal transport through two broad mechanisms: ballistic displacement damage, which creates native defects such as vacancies, interstitials, antisites, defect clusters, and extended defects, and neutron-induced transmutation, which changes the local chemical composition by introducing impurities. The relative importance of these mechanisms depends strongly on the neutron energy spectrum, fluence, dose rate, and irradiation temperature. Fission reactor spectra can produce substantial displacement damage in SiC-based fuel and cladding components, whereas the harder high-energy neutron spectrum in fusion environments generally enhances high-threshold gas production

* These authors contributed equally to this work.
† mppolak@wisc.edu
‡ szlufarska@wisc.edu

and metallic transmutation reactions [10, 11].

Native defects strongly scatter phonons and can cause a severe loss of thermal conductivity, regardless of the initial material quality. Experiments based on Raman spectroscopy and time-domain thermoreflectance have shown that the thermal conductivity of high-purity CVD SiC can drop from several hundred $\mathrm{W\,m^{-1}\,K^{-1}}$ to only a few $\mathrm{W\,m^{-1}\,K^{-1}}$ after irradiation, with the degradation often saturating at doses of about 1-5 dpa when the irradiation temperature is below 773 K [3, 12–15]. Sintered SiC, which already contains more intrinsic defects and has a lower baseline conductivity, shows a similar trend under comparable irradiation conditions [12]. In addition to displacement damage, neutron-induced transmutation can generate H, He, Be, Mg, Al, and P in SiC, where the production rates and dominant species are spectrum dependent [16–18]. In fusion-relevant high-energy neutron spectra, Mg has been predicted to be the dominant solid transmutant in SiC, whereas meaningful Mg production is not expected under typical fission-reactor irradiation because the neutron energies are generally insufficient for the relevant reactions [17, 18]. In realistic irradiation conditions, native defects and transmutation products can coexist and interact, and their combined presence can further modify phonon transport. However, the coupled effect of intrinsic irradiation defects and neutron-induced Mg transmutation on thermal transport in 3C-SiC has not yet been quantified.

For SiC, thermal transport is dominated by the lattice thermal conductivity and is thus primarily governed by phonon transport. As a typical wide-bandgap covalent compound, SiC exhibits an electronic contribution to heat conduction that is generally much smaller than the lattice contribution. Previous studies on SiC-based ceramics [19–21], including composite, porous, and liquid-phase-sintered systems, have reported electrical conductivities ranging from approximately $10^3$ S/m to $10^{-2}$ S/m. Using the Wiedemann–Franz relation as an order-of-magnitude estimate, the corresponding electronic thermal conductivity is on the order of $10^{-3}$ to $10^{-8}$ $\mathrm{W\,m^{-1}\,K^{-1}}$, which is significantly smaller than the lattice thermal conductivity of SiC (about 450 $\mathrm{W\,m^{-1}\,K^{-1}}$). Similarly, irradiated SiC [22] has been reported to exhibit an electrical conductivity of approximate 1.4 S/m, corresponding to an electronic thermal conductivity of only about $10^{-5}$ $\mathrm{W\,m^{-1}\,K^{-1}}$. These results indicate that heat conduction in SiC is overwhelmingly dominated by phonons, and that elucidating the microstructure–phonon transport relationship is therefore critical for understanding the influence of defects on its thermal properties.

Experimental measurements of thermal conductivity are often affected by heat loss, nonuniform heating, and sample-to-sample variations in microstructure. As a result, reported thermal conductivities for pristine 3C-SiC span a wide range, typically from about 300 to 520 $\mathrm{W\,m^{-1}\,K^{-1}}$, depending on grain size, residual defects, dopants, and processing history [23–25]. Computational approaches, therefore, provide an important complement, because they offer controlled model systems and direct access to phonon-scattering mechanisms. Substantial progress has been made in understanding thermal transport in 3C-SiC using both lattice-dynamics (LD) calculations based on the Boltzmann transport equation (BTE) and molecular-dynamics (MD) simulations. BTE-based approaches rely on phonon frequencies, group velocities, and scattering rates derived from atomistic force constants, whereas MD methods include nonequilibrium, approach-to-equilibrium, and equilibrium formulations such as the Green-Kubo method. In all cases, predictive accuracy depends critically on the underlying description of interatomic interactions.

Interatomic interactions can be described either from first principles, most commonly with density functional theory (DFT), or with interatomic potentials. DFT provides high accuracy, but its computational cost limits applications to relatively small and highly symmetric systems. Classical empirical interatomic potentials allow simulations on much larger length and time scales, but they often suffer from limited accuracy and poor transferability outside the fitting space. This limitation is especially problematic for thermal transport. Quantitative prediction of thermal conductivity requires accurate second- and third-order derivatives of the energy with respect to atomic displacements. These quantities control harmonic phonons and anharmonic phonon scattering, particularly the three-phonon processes governed by third-order force constants. Conventional empirical potentials can struggle to reproduce these quantities with sufficient fidelity, which limits their predictive power for thermal conductivity.

Machine-learning interatomic potentials (MLIPs) provide a promising alternative by combining near-DFT accuracy with the efficiency needed for large-scale simulations [26, 27]. When trained on diverse and well-converged DFT datasets, MLIPs can describe both harmonic and anharmonic interactions with substantially improved accuracy and transferability compared to classical interatomic potential. Several MLIPs have already been developed for SiC and used for radiation and/or thermal transport modeling. For example, Fu *et al.* used a deep-learning potential to study phonons and thermal conductivity in SiC with near-quantum accuracy [28]. Zhang *et al.* developed an MLIP for thermal transport in SiC with different polytypes and stacking fault types in the 4H-SiC [29]. Liu *et al.* constructed a neural-network potential to investigate defect formation induced by knock-on irradiation in 4H-SiC [30]. However, most previous studies have focused either on pristine SiC and its polytypes or on intrinsic defects alone. To our knowledge, the combined effect of intrinsic irradiation defects and Mg transmutation on thermal transport in 3C-SiC has not been addressed.

In this work, we develop an MLIP for SiC containing both intrinsic defects and Mg-related defects, denoted MLIP4SiC-Mg. The model is trained on a com-prehensive DFT dataset comprising multiple SiC polytypes and a broad range of intrinsic and Mg-containing defect configurations. Benchmark calculations show that MLIP4SiC-Mg reproduces DFT energies, forces, phonon dispersions, and lattice thermal conductivities with near-DFT accuracy. Using this validated potential, we then examine the thermal transport behavior of intrinsic point defects, Mg-related point defects, and small defect complexes by equilibrium molecular dynamics (EMD). To improve the quantitative reliability of the thermal conductivity, we further employ a correction scheme based on controlled Langevin noise and extrapolation to the zero-force-error limit [31]. For defective systems in the dilute-defect regime, this framework is extended through a resistance-based correction treatment. The results reveal a strong and configuration-dependent effect of Mg-related defects on thermal transport in 3C-SiC and provide an efficient atomistic framework for studying irradiation-induced thermal conductivity degradation in SiC.

## II. METHODS

### A. Density functional theory

DFT calculations were performed using the Vienna *Ab initio* Simulation Package (VASP) within the Perdew-Burke-Ernzerhof (PBE) generalized gradient approximation and the projector augmented wave (PAW) method [32–34]. For structural relaxations, a plane-wave energy cutoff of 600 eV was used. Brillouin-zone integrations were carried out using Monkhorst-Pack $k$-point meshes, including $4 \times 4 \times 3$ and $2 \times 2 \times 3$ grids depending on the supercell size. The electronic and ionic convergence criteria were set to $10^{-5}$ eV and $10^{-3}$ eV/Å, respectively. For static self-consistent field calculations, the plane-wave energy cutoff was increased to 800 eV, and a $3 \times 4 \times 4$ Monkhorst-Pack $k$-point mesh was used for the $2 \times 2 \times 3$ defect-free 3C-SiC supercell containing 96 atoms. The electronic energy convergence criterion for these calculations was set to $10^{-7}$ eV.

### B. Construction of the machine-learning interatomic potential

The neuroevolution potential (NEP) was developed using the Graphics Processing Units Molecular Dynamics (GPUMD) package [35]. NEP is a neural-network-based machine-learning potential trained using the separable natural evolution strategy (SNES), and has been widely used in atomistic simulations of thermal transport [36]. Following the Behler-Parrinello high-dimensional neural-network framework, the site energy of atom $i$ is expressed as a function of a local descriptor vector with $N_{\mathrm{des}}$ components, $U_i(q) = U_i\left(\{q_v^i\}_{v=1}^{N_{\mathrm{des}}}\right)$. In the NEP formalism, the site energy is represented by a feedforward neural network with a single hidden layer containing $N_{\mathrm{neu}}$ neurons:

$$U_i = \sum_{\mu=1}^{N_{\mathrm{neu}}} \omega_\mu^{(1)} \tanh\left(\sum_{v=1}^{N_{\mathrm{des}}} \omega_{\mu v}^{(0)} q_v^i - b_\mu^{(0)}\right) - b^{(1)}. \quad (1)$$

Here, $\tanh(x)$ is the activation function, and $\omega^{(0)}$, $\omega^{(1)}$, $b^{(0)}$, and $b^{(1)}$ are the trainable weights and biases of the neural network. In NEP, the local descriptor $q_v^i$ consists of radial and angular components. The radial terms depend only on pair distances and are constructed from Chebyshev polynomials, whereas the angular terms incorporate angular information through spherical harmonics in a form related to the atomic cluster expansion. The training hyperparameters are listed in Table S1.

Training structures were generated by MD simulations using the Large-scale Atomic/Molecular Massively Parallel Simulator (LAMMPS) package [37]. Pristine SiC and defective structures were sampled using the NEP89 potential [38]. The MD simulations were performed in the NVT or NPT ensemble with a Nosé-Hoover thermostat, using a time step of 1 fs and total sampling simulation times of 1.7 ns and 1.6 ns, respectively. To sample a broad range of configurations, the simulations included both heating runs over the temperature ranges of 2000-5000 K, 5000-6000 K, and 300-6000 K, and constant-temperature runs at 300, 600, 900, and 1200 K. For each temperature interval, 500 structures were sampled, as summarized in Fig. S1.

### C. MD simulations of thermal conductivity

Equilibrium molecular dynamics (EMD) simulations combined with the Green-Kubo formalism were performed in GPUMD to evaluate the thermal conductivity [35]. In this approach, the thermal conductivity is obtained from the time integral of the heat-current autocorrelation function [39]:

$$\kappa_\alpha = \frac{1}{k_B T^2 V} \int_0^{t_0} \langle J_\alpha(0) J_\alpha(t) \rangle \, dt, \qquad \alpha = x, y, z \quad (2)$$

where $k_B$ is the Boltzmann constant, $V$ is the system volume, $T$ is the temperature, and $t_0$ is the upper limit of the correlation time integral. $J_\alpha$ is the $\alpha$ component of the heat current $\mathbf{J}$, which is given by [39]

$$\mathbf{J} = \sum_i \mathbf{v}_i E_i + \sum_i \sum_{j \neq i} \mathbf{r}_{ij} \left(\frac{\partial U_j}{\partial r_{ij}} \mathbf{v}_i\right), \quad (3)$$

Here, $\mathbf{v}_i$ and $E_i$ are the velocity and total energy of atom $i$, respectively, $\mathbf{r}_{ij}$ is the position vector between atoms $i$ and $j$, and $U_j$ is the potential energy of atom $j$.

A 13,824-atom supercell was used for all thermal-conductivity calculations. Defective structures were generated by introducing vacancies or substitutional defects at randomly selected lattice sites in the pristine 3C-SiC supercell. Each simulation first employed the NVT ensemble for 200 ps to equilibrate the system at the target temperature, followed by a 15 ns production run in the NVE ensemble for heat-current sampling. Periodic boundary conditions were applied in all three directions, and the time step was set to 1 fs. A correlation time of 150 ps was sufficient for the thermal conductivity to converge. For each condition, 10 independent simulations were performed, and the reported thermal conductivity and error bars were obtained by averaging and taking the standard deviation over these runs, respectively.

Classical MD, like that performed in this work, neglects Bose-Einstein phonon statistics, so it is only an approximation to the true quantum dynamics. This approximation is usually accurate when the temperature is above the Debye temperature, denoted $T_D$. The Debye temperature can be approximately thought of as the temperature scale associated with the highest phonon frequencies in a solid, with the corresponding energy given by $k_B T_D \approx \hbar\omega_D$, where $\omega_D$ is the Debye frequency. For 3C-SiC, $T_D$ is about 1200 K.

Below $T_D$, classical MD can become less accurate because phonon occupations and modal heat capacities are treated classically rather than quantum mechanically. However, the relevant temperature scale is not necessarily the full Debye temperature of all phonon modes. For thermal conductivity, what matters most is the energy scale of the phonons that carry most of the heat, which can be thought of as having an effective Debye temperature, $T_{D,\text{eff}}$, above which they can be accurately treated as classical. In SiC, these dominant heat carriers are mainly long-lived acoustic phonons, which leads to $T_{D,\text{eff}} \ll T_D$.

This point helps explain why classical MD is still widely used for SiC thermal-conductivity calculations below $T_D$. Such calculations are common for pristine and defective systems, including bulk polytypes, nanostructures, and systems with point defects and irradiation damage [40–45]. The use of classical MD below $T_D$ is at least partly justified because thermal transport in SiC is dominated by acoustic branches. The low-lying optical branches have relatively small group velocities. They mainly reduce $\kappa$ by adding scattering channels for acoustic phonons, rather than by carrying a large fraction of the heat themselves [46]. Therefore, the total thermal conductivity $\kappa$ is less sensitive to the inaccurate classical occupation of high-frequency optical modes. In Fig. 6 we compare thermal conductivity for pure SiC obtained with MLIP4SiC-Mg in classical MD as applied in this work (Corrected MLMD), and with a fully quantum mechanical BTE treatment (MLIP4SiC-Mg (BTE)). The results match each other, and fit well within reference experimental values, further strengthening the applicability of classical MD, to thermal conductivity of SiC.

It is also plausible that classical MD below $T_D$ but above $T_{D,\text{eff}}$ can still effectively capture how defects reduce $\kappa$, because it can capture how defects perturb the lifetimes of the acoustic phonons that dominate heat transport. These lifetime reductions can arise from mass disorder, force-constant disorder, and local lattice distortion [45]. Previous classical MD studies have shown that vacancies, antisites, interstitials, microvoids, and cascade damage can strongly reduce $\kappa$ in SiC [40, 42–44]. First-principles calculations that use a quantum treatment of transport support this picture of defect scattering in SiC. They show that some substitutional impurities can cause very strong, and even resonant, phonon-defect scattering. They also show that the very high $\kappa$ of 3C-SiC is recovered only when defect scattering is strongly suppressed [47, 48].

Separate from the roles of different phonon contributions, there are fortuitous error cancellations in the classical approximations to the full quantum behavior that can allow classical MD to give reasonable total $\kappa$ values below $T_D$. Classical statistics overestimate modal heat capacities and phonon populations. However, they also increase phonon-phonon scattering and shorten phonon lifetimes. These two errors go in opposite directions and are expected to at least partially cancel. As a result, the total thermal conductivity predicted by classical MD can be closer to experimental values than either error alone would suggest [49–52].

Given the above observations, we follow the common practice in the field of using classical MD for our SiC studies, even well below $T_D$. However, it is important to recognize that this is an approximation and that some errors are to be expected. Quantifying the magnitude of these classical-approximation errors in the present work and related studies remains an important topic for future investigation.

## III. RESULTS AND DISCUSSION

### A. Workflow and database for SiC with intrinsic defects and Mg transmutants

To describe thermal transport in defective 3C-SiC, the interatomic potential must remain accurate across a wide range of local environments, including pristine crystal structures, intrinsic defects, Mg-related defects, and their combinations. This requirement is particularly important for phonon dispersion and lattice thermal conductivity, both of which are highly sensitive to the quality of the predicted forces. To this end, we constructed a comprehensive DFT dataset for SiC containing intrinsic defects and Mg transmutants, and trained the potential within the NEP framework [35]. The overall workflow for database construction, model training, and property evaluation is summarized in Fig. 1.

The final database contains 47,533 configurations and 4,563,168 atomic environments. For each configuration,

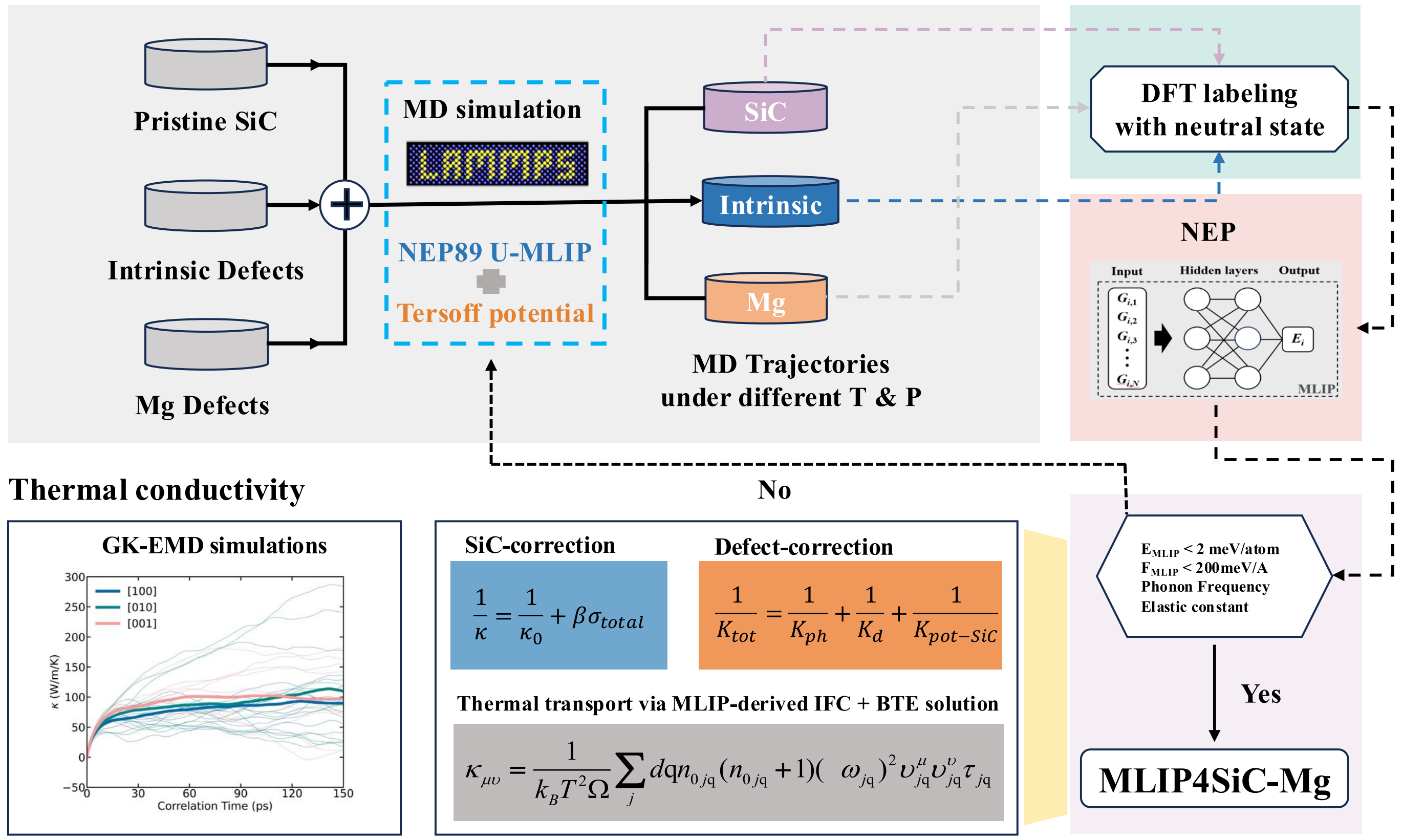

FIG. 1. Workflow for database construction, NEP training, and property evaluation.

DFT total energies, atomic forces, and virial stresses were collected to train MLIP4SiC-Mg. As shown in Fig. 2, the dataset is organized into three main groups: (i) pristine crystalline SiC and related SiC polymorphs; (ii) intrinsic defects, including vacancies, substitutional defects, and interstitials; and (iii) Mg-related defects, including point defects such as $Mg_{Si}$, $Mg_{C}$, $Mg_{TC}$, and $Mg_{TSi}$, as well as Mg-containing complexes such as $Mg_{Si}$-$V_{C}$ and $Mg_{Si}$-$V_{Si}$. For the pristine and polymorph portion of the dataset, we also incorporated previously reported SiC databases from Refs. [29, 53]. In addition, 800 computed Si-C-Mg structures from the Materials Project database were included to further expand the structural diversity [54]. The number of configurations used for different structures is summarized in Table S2.

For each defect category, the training structures were generated using an NVT+NPT sampling strategy. Specifically, equilibrium NVT simulations at selected temperatures were followed by NPT simulations over a range of temperatures, as described in the Methods section. The low-temperature sampling is important for accurately representing crystalline SiC and defect energetics near equilibrium. The higher-temperature sampling introduces distorted, low-symmetry, and high-energy configurations, thereby broadening the configurational space covered by the training set. This combination of near-equilibrium and strongly perturbed structures improves the transferability of MLIP4SiC-Mg across the defect environments relevant to thermal-transport calculations.

### B. Training and testing results

The predictive accuracy of MLIP4SiC-Mg was first evaluated using parity plots for energies and forces, as shown in Fig. 3a-b and Fig. 3e-f. The root mean square error (RMSE) was used to quantify the agreement between the MLIP predictions and the DFT reference data. For both the training and testing sets, the energy RMSE is approximately 10 meV/atom, while the force RMSEs are 297.8 and 295.2 meV/Å, respectively. These errors are comparable to those reported for NEP-based machine-learning interatomic potentials trained on structurally complex systems containing diverse local environments, such as defective, amorphous, or liquid-like configurations [55–57]. The close agreement between training and testing errors also indicates that the model does not show obvious overfitting. Overall, these results show that MLIP4SiC-Mg provides a reliable description of the DFT reference dataset across the broad range of pristine, defective, and Mg-containing SiC configurations considered in this work.

The error distributions for the testing set are shown in Fig. 3c-d and Fig. 3g-h. Both the energy and force residuals are centered close to zero, indicating that MLIP4SiC-

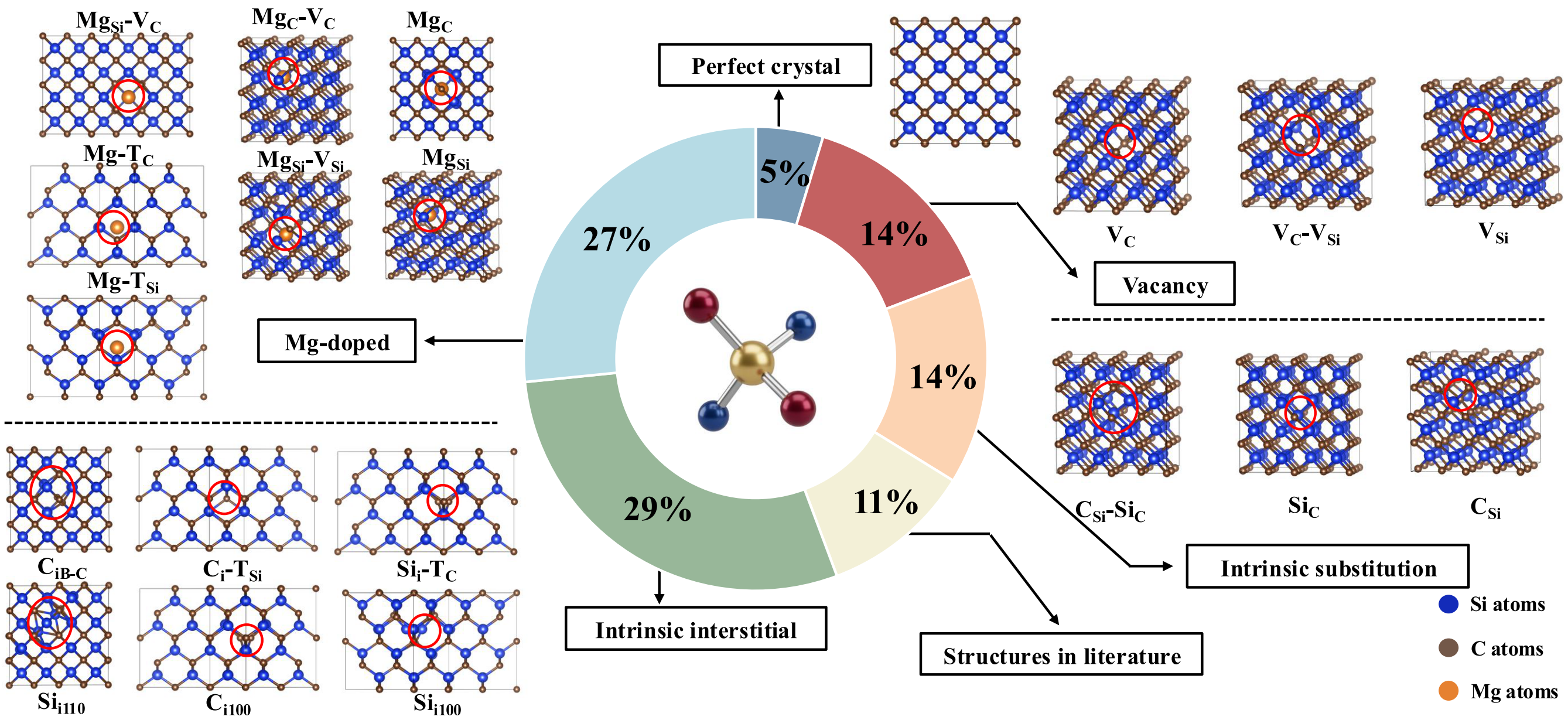

FIG. 2. Overview of the DFT training database used to construct MLIP4SiC-Mg. The dataset includes pristine SiC and related polymorphs, intrinsic defects on both C and Si sublattices, isolated Mg defects, and selected short-range Mg-defect complexes. The central donut chart shows the fraction of configurations in each major structure class, while the surrounding panels show representative atomic configurations. Detailed numbers of configurations for each defect type are summarized in Table S2.

Mg does not exhibit a significant systematic bias over the testing configurations. Additionally, the fact that most residuals are concentrated near zero shows that the model accurately describes the majority of local environments in the testing set. A small fraction of larger residuals is observed, particularly for the force components, which is expected for a heterogeneous training database containing highly distorted, high-energy, and complex defect-containing configurations. These more challenging configurations increase the diversity of the database and help improve the robustness of the potential beyond near-equilibrium crystalline environments. Overall, the small testing errors, the near-zero-centered residuals, and the successful reproduction of phonon dispersions and thermal conductivities indicate that MLIP4SiC-Mg provides a reliable description of both pristine and defective SiC configurations relevant to the present thermal-transport calculations.

### C. Performance of the MLIP4SiC-Mg potential

We next examined the transferability of MLIP4SiC-Mg by comparing its equation of state predictions for representative pristine and defective structures, including 3C-SiC, $V_{Si}$, and $Mg_{Si}$. The structural properties are summarized in Table I. For pristine 3C-SiC, the MLIP4SiC-Mg results agree closely with DFT, with the largest deviation among the listed structural parameters being only 0.35%. The predicted lattice constant is also close to the experimental value [58, 59].

The same level of agreement is retained for the defective systems, as shown in Fig. 4. MLIP4SiC-Mg reproduces the defect-induced lattice expansion and the associated reduction in bulk modulus for both $V_{Si}$ and $Mg_{Si}$. The reasonable agreement in $B'_0$ further indicates that the model captures the third derivatives of the energy-volume relation. Together, these results support that MLIP4SiC-Mg is sufficiently transferable to describe the energetics and structural response of both pristine and defective SiC.

To further assess the ability of MLIP4SiC-Mg to describe vibrational properties, we calculated the phonon dispersion curves of 3C-SiC, $V_{Si}$, and $Mg_{Si}$ within the harmonic approximation and compared them with the DFT results. As shown in Fig. 5a-c, the MLIP4SiC-Mg phonon spectra agree well with DFT for all three systems. For pristine 3C-SiC, the agreement is excellent over the full frequency range. For $V_{Si}$ and $Mg_{Si}$, the model also reproduces the main features of the defect-perturbed phonon bands, including band splitting, discontinuities, and broadening. Small deviations appear in the highest optical modes, but these differences are not expected to significantly affect the overall thermal transport behavior. These results confirm that MLIP4SiC-Mg provides a reliable description of the harmonic vibrational properties of both pristine and defective SiC.

Beyond the harmonic phonon properties, we evaluated the anharmonic thermal transport predicted by MLIP4SiC-Mg using lattice-dynamics calculations based on the Boltzmann transport equation (BTE). The resulting lattice thermal conductivities were directly compared with DFT-BTE results. Because the calculated thermal conductivity is sensitive to the third-order interatomic force constant cutoff, the $q$-point sampling density, and the equilibrium lattice parameters, identical computational settings were used for the MLIP4SiC-Mg-BTE and DFT-BTE calculations to ensure a direct comparison. The calculation workflow is summarized in Fig. S2.

TABLE I. Structural properties of 3C-SiC, $V_{Si}$, and $Mg_{Si}$: total energy $E_0$ (in eV/atom), lattice constant $a_0$ (in Å), bulk modulus $B_0$ (in GPa), and its pressure derivative $B_0'$.

| | **SiC** | | | **$V_{Si}$** | | **$Mg_{Si}$** | |
|---|---|---|---|---|---|---|---|
| Property | MLIP4SiC-Mg | DFT | EXP | MLIP4SiC-Mg | DFT | MLIP4SiC-Mg | DFT |
| $E_0$ | -7.533 | -7.521 | – | -7.443 | -7.457 | -7.403 | -7.406 |
| $a_0$ | 4.379 | 4.376 | 4.360 [58] | 8.769 | 8.759 | 8.791 | 8.780 |
| $B_0$ | 264.692 | 273.531 | 260.100 [59] | 32.692 | 33.871 | 32.24 | 33.770 |
| $B_0'$ | 3.930 | 3.805 | 4.000 [60] | 3.955 | 4.719 | 3.939 | 4.556 |

The predicted average thermal conductivities of pristine and defective SiC are shown in Fig. 5d-f. MLIP4SiC-Mg reproduces the DFT-BTE results well over the entire temperature range from 200 to 1000 K. At 300 K, the thermal conductivity of pristine 3C-SiC predicted by MLIP4SiC-Mg is 498 $Wm^{-1}K^{-1}$, in excellent agreement with the DFT value of 500 $Wm^{-1}K^{-1}$ when only three-phonon scattering is included. This value is also consistent with reported experimental data for high-quality single-crystal 3C-SiC [23]. For the defective systems, both the reduction trends and the absolute values of $\kappa$ predicted by MLIP4SiC-Mg closely follow the DFT results, further supporting the ability of the potential to capture defect-induced phonon scattering.

It is also worth noting that the predicted thermal conductivity of pristine 3C-SiC falls within the experimentally reported range of about 300-520 $W\,m^{-1}\,K^{-1}$ [23–25]. The spread in the experimental data is expected, since measured values depend on crystal quality, grain size, residual defects, dopants, and measurement conditions. The overall agreement of MLIP4SiC-Mg with both DFT and experiment demonstrates that the model can accurately describe not only intrinsic phonon thermal transport in SiC, but also the changes introduced by representative defect structures [61].

### D. Thermal conductivity modeling of irradiation-induced defects in SiC using MLIP-MD

Under irradiation, SiC typically accumulates a variety of defects, including intrinsic point defects, Mg transmutation products, and their coupled complexes. The concentrations of these defects depend on irradiation dose, temperature, reactor conditions, and post-irradiation annealing history [61]. Evaluating thermal conductivity across such a broad defect space generally requires large supercells to capture both defect-phonon scattering and concentration effects. In practice, however, these systems are often beyond the feasible size range of lattice-dynamics calculations based on the Boltzmann transport equation (LD+BTE), because the calculation of third-order interatomic force constants becomes prohibitively expensive for supercells containing thousands of atoms.

To address this limitation, we employed MD simulations with the newly developed MLIP. Specifically, thermal conductivity was evaluated using EMD together with the Green-Kubo relation, which enables efficient calculations over a wide range of defect types, concentrations, and temperatures. Based on the defect stability analysis in our previous work [62], and consistent with defect energetics reported in the literature [63], we focus on four representative defects in 3C-SiC: $V_C$, $Mg_{Si}$, $Mg_{TC}$, and the Mg-containing defect complex $Mg_{Si}$-$V_C$. For each defect type, thermal conductivity was evaluated as a function of both temperature and defect concentration. In the present work, two representative defect concentrations and five temperatures were considered. We used a method of randomly deleting or replacing a certain number of atoms to generate defect structures. To facilitate comparison of thermal conductivity under the same defect concentration, two models are added for $Mg_{Si}$-$V_C$ to maintain uniformity in concentration.

For materials with intrinsically high lattice thermal conductivity, including $CoSb_3$ [64] and cubic Si [65], MLIP-based MD simulations have been shown to systematically underestimate $\kappa$ relative to experiment. A similar trend was also reported for 3C-SiC using the NEP potential with the HNEMD method [29]. To obtain quantitatively reliable thermal conductivities, we therefore adopted the correction scheme proposed by Wu *et al.* [31]. In this approach, controlled stochastic force noise is introduced through a Langevin thermostat, and the resulting thermal conductivities are extrapolated to the zero-force-error limit:

$$\frac{1}{\kappa} = \frac{1}{\kappa_0} + \beta\sigma_{\text{total}}. \quad (4)$$

Here, $\kappa$ is the thermal conductivity obtained from MLIP-based MD simulations in the presence of a total force-error variance $\sigma_{\text{total}}$, $\kappa_0$ is the extrapolated thermal conductivity at zero force error, and $\beta$ is a fitting parameter. The physical basis of this method is that both the intrinsic MLIP force errors and the random forces generated by the Langevin thermostat approximately follow Gaussian distributions. By introducing Langevin noise with known variance, evaluating $\kappa$ at several noise levels, and extrapolating to $\sigma_{\text{total}} = 0$, one can estimate the intrinsic thermal conductivity in the absence of force errors. The total force variance is defined as [31]

$$\sigma_{\text{total}}^2 = \sigma_L^2 + \sigma_{\text{mlp}}^2, \tag{5}$$

where $\sigma_{\text{mlp}}$ denotes the standard deviation of the intrinsic MLIP force error, and $\sigma_L$ denotes the standard deviation of the random force introduced by the Langevin thermostat. For the pristine SiC correction, $\sigma_{\text{mlp}}$ was estimated using representative 300 K pristine configurations extracted from active-learning MD trajectories and relabeled by DFT. The resulting force-error standard deviation between MLIP4SiC-Mg and the DFT reference is $\sigma_{\text{mlp}} = 26.93\ \text{meV}/\text{Å}$. This value is substantially smaller than the global testing force RMSE because it corresponds specifically to near-equilibrium pristine SiC, rather than to the full heterogeneous testing set containing distorted and defect-containing configurations. The Langevin force variance is given by [31]

$$\sigma_L^2 = \frac{2k_B T m}{\tau_T \Delta t}, \tag{6}$$

where $m$ is the average atomic mass of the system, $k_B$ is the Boltzmann constant, $\Delta t$ is the time step, and $\tau_T$ is the thermostat relaxation time. In this work, $\tau_T$ values of 40, 250, and 450 ps were used to generate different noise levels for the extrapolation; additional details are provided in the Supporting Information. Using this procedure, we obtained a corrected thermal conductivity of 421 $\text{W}\,\text{m}^{-1}\,\text{K}^{-1}$ for pristine 3C-SiC at 300 K, in good agreement with experiment, as shown in Fig. 6. A small difference from experiment is expected because experimental samples may contain vacancies, dislocations, and other residual defects, and may also be affected by finite-size and boundary scattering [66, 67].

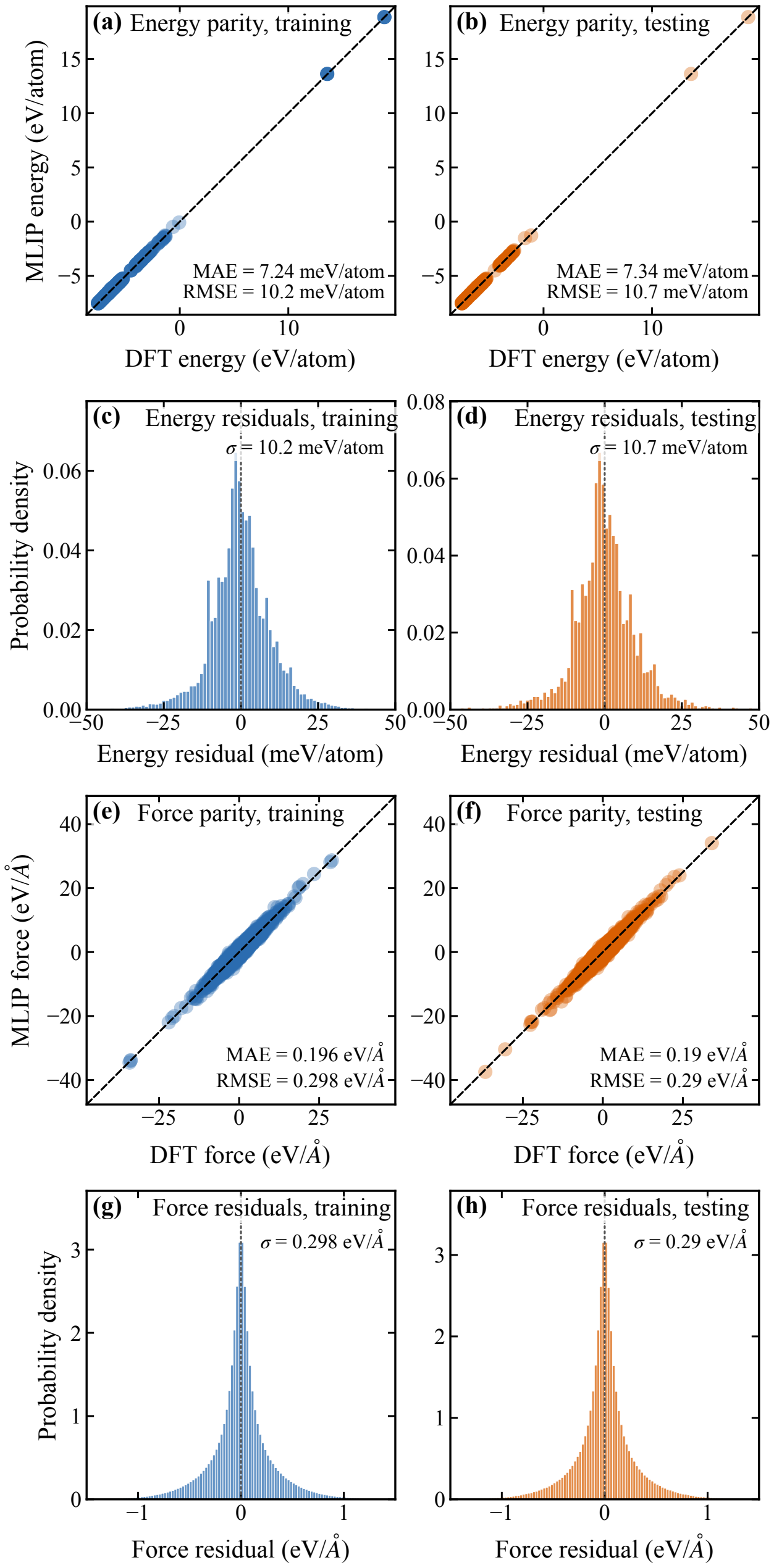

FIG. 3. Parity plots comparing DFT reference energies and forces with MLIP4SiC-Mg predictions for the training and testing datasets. The lower panels show the corresponding residual distributions for the testing set, indicating that most energy and force residuals are concentrated near zero with no significant systematic bias.

For defective systems, direct Langevin-thermostat extrapolation is less robust than for pristine 3C-SiC, because defect-induced scattering broadens the heat-current correlation function and makes the noise-level extrapolation less stable. In addition, for dilute defects, most of the MLIP-induced correction originates from the host-like SiC environment rather than from the small fraction of atoms near the defect. We therefore apply the correction in thermal-resistance space using an approximate Matthiessen-type treatment. First, we define the thermal resistance as

$$R \equiv \kappa^{-1}. \tag{7}$$

For a defective system calculated with MLIP-based MD, the total resistance can be written as

$$R_{\text{tot}}^{\text{ML}} = R_{\text{ph}} + R_{\text{def}} + R_{\text{pot}}^{\text{def}}, \tag{8}$$

where $R_{\text{ph}}$ is the intrinsic phonon-phonon resistance of the host lattice, $R_{\text{def}}$ is the physical defect-induced resistance, and $R_{\text{pot}}^{\text{def}}$ is the additional nonphysical resistance associated with MLIP force errors in the defective system. The corrected resistance we seek is the physical resistance of the defective system after removing this nonphysical MLIP contribution:

$$R_{\mathrm{tot}}^{\mathrm{corr}} \equiv R_{\mathrm{ph}} + R_{\mathrm{def}}. \tag{9}$$

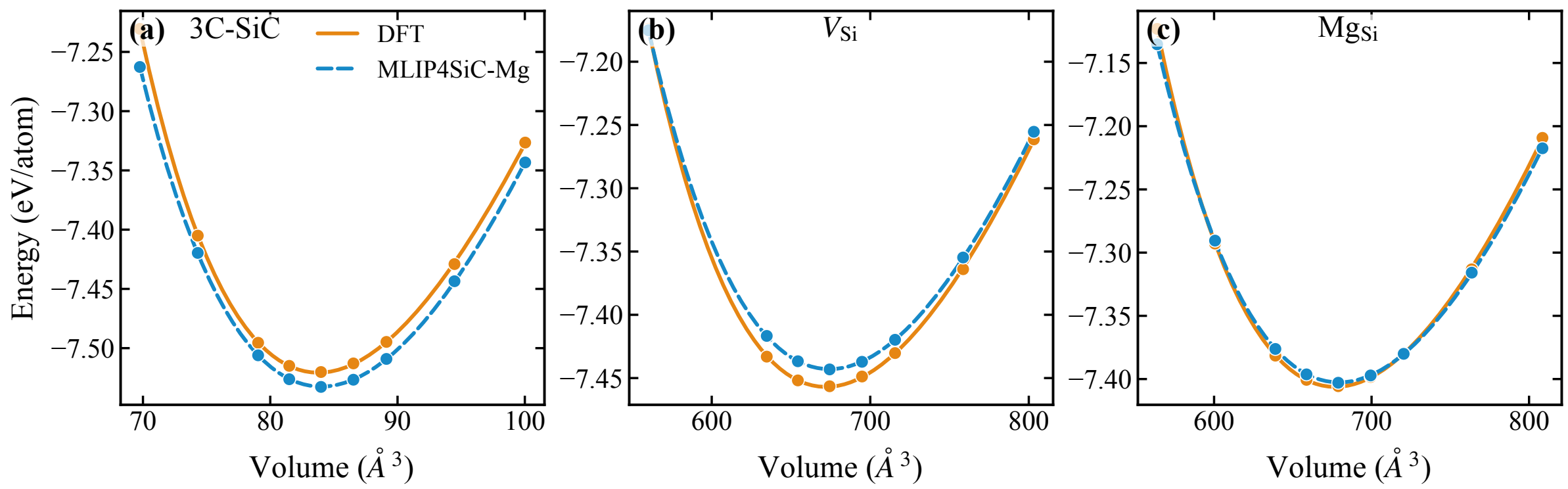

FIG. 4. Energy-volume relations for 3C-SiC, $V_{Si}$, and $Mg_{Si}$ calculated using DFT and MLIP4SiC-Mg. The data were fitted using the Murnaghan equation of state, with the fitted parameters presented in Tab. I.

Combining Eqs. (8) and (9) gives

$$R_{\mathrm{tot}}^{\mathrm{corr}} = R_{\mathrm{tot}}^{\mathrm{ML}} - R_{\mathrm{pot}}^{\mathrm{def}}. \tag{10}$$

The remaining task is to estimate $R_{\mathrm{pot}}^{\mathrm{def}}$. For a dilute defect concentration $c$, the MLIP-induced resistance in the defective supercell can be written as a pristine host contribution plus a defect-dependent correction:

$$R_{\mathrm{pot}}^{\mathrm{def}} = R_{\mathrm{pot}}^{\mathrm{SiC}} + \Delta R_{\mathrm{pot}}^{\mathrm{def}}, \tag{11}$$

where $\Delta R_{\mathrm{pot}}^{\mathrm{def}}$ arises from the small fraction of defect-centered local environments. Because all defect concentrations considered here are in the dilute regime, $\Delta R_{\mathrm{pot}}^{\mathrm{def}}$ is expected to be a higher-order correction relative to the dominant host-like MLIP contribution. We therefore approximate

$$R_{\mathrm{pot}}^{\mathrm{def}} \approx R_{\mathrm{pot}}^{\mathrm{SiC}}. \tag{12}$$

Substituting Eq. (12) into Eq. (10) yields

$$R_{\mathrm{tot}}^{\mathrm{corr}} \approx R_{\mathrm{tot}}^{\mathrm{ML}} - R_{\mathrm{pot}}^{\mathrm{SiC}}. \tag{13}$$

The pristine MLIP-induced resistance can be obtained from the difference between the uncorrected and corrected pristine SiC resistances:

$$R_{\mathrm{pot}}^{\mathrm{SiC}} = R_{\mathrm{SiC}}^{\mathrm{ML}} - R_{\mathrm{SiC}}^{\mathrm{corr}}. \tag{14}$$

Therefore, the corrected resistance of the defective system is estimated as

$$R_{\mathrm{tot}}^{\mathrm{corr}} \approx R_{\mathrm{tot}}^{\mathrm{ML}} - \left(R_{\mathrm{SiC}}^{\mathrm{ML}} - R_{\mathrm{SiC}}^{\mathrm{corr}}\right). \tag{15}$$

Only after this resistance-level correction is obtained do we convert back to thermal conductivity:

$$\left(\kappa_{\mathrm{def}}^{\mathrm{corr}}\right)^{-1} \approx \left(\kappa_{\mathrm{def}}^{\mathrm{ML}}\right)^{-1} - \left[\left(\kappa_{\mathrm{SiC}}^{\mathrm{ML}}\right)^{-1} - \left(\kappa_{\mathrm{SiC}}^{\mathrm{corr}}\right)^{-1}\right]. \tag{16}$$

This approximation preserves the physical defect scattering contained in the MLIP-MD result while removing the dominant host-like MLIP force-error contribution. Its validity relies on the dilute-defect assumption that the defect-dependent variation in the MLIP correction is small compared with the pristine host contribution.

Using this resistance-based correction scheme, we evaluated the thermal conductivity of defective 3C-SiC over the temperature range considered. All four defect types lead to a pronounced reduction in thermal conductivity relative to pristine 3C-SiC. At 300 K, the corrected thermal conductivity of pristine 3C-SiC is about 421 $\mathrm{W\,m^{-1}\,K^{-1}}$, whereas it decreases to ∼128 and ∼60 $\mathrm{W\,m^{-1}\,K^{-1}}$ for $V_C$ at 0.065 and 0.260 at.%, respectively, to ∼180 and ∼79 $\mathrm{W\,m^{-1}\,K^{-1}}$ for $Mg_{Si}$ at the same concentrations, and to ∼123 and ∼59 $\mathrm{W\,m^{-1}\,K^{-1}}$ for $Mg_{TC}$. For the $Mg_{Si}$-$V_C$ complex, the thermal conductivity at 300 K decreases from ∼140 $\mathrm{W\,m^{-1}\,K^{-1}}$ at 0.072 at.% to ∼81 $\mathrm{W\,m^{-1}\,K^{-1}}$ at 0.521 at.%. Furthermore, taking $Mg_{Si}$ as an example, two methods (the one proposed by Wu *et al.* and the one in this work) were used to correct the thermal conductivity of the defect structure. The deviation between the two methods was small, which shows that our proposed method is effective, as shown in Table S3.

These results show that even dilute defects can strongly disrupt heat transport in 3C-SiC. The suppression is most pronounced at low temperature, where heat conduction is dominated by long-mean-free-path phonons that are highly sensitive to mass disorder, force-constant perturbations, and local lattice distortion. As the temperature increases, the thermal conductivity of all systems decreases monotonically, while the differences among defect types become progressively smaller. At 1500 K, the thermal conductivities of all defective systems fall within a relatively narrow range of about 25-39 $\mathrm{W\,m^{-1}\,K^{-1}}$, much smaller than the spread at 300 K. This trend is expected, because intrinsic anharmonic phonon-phonon scattering becomes increasingly important at high temperature and reduces the relative contri-bution of defect scattering to the total thermal resistance.

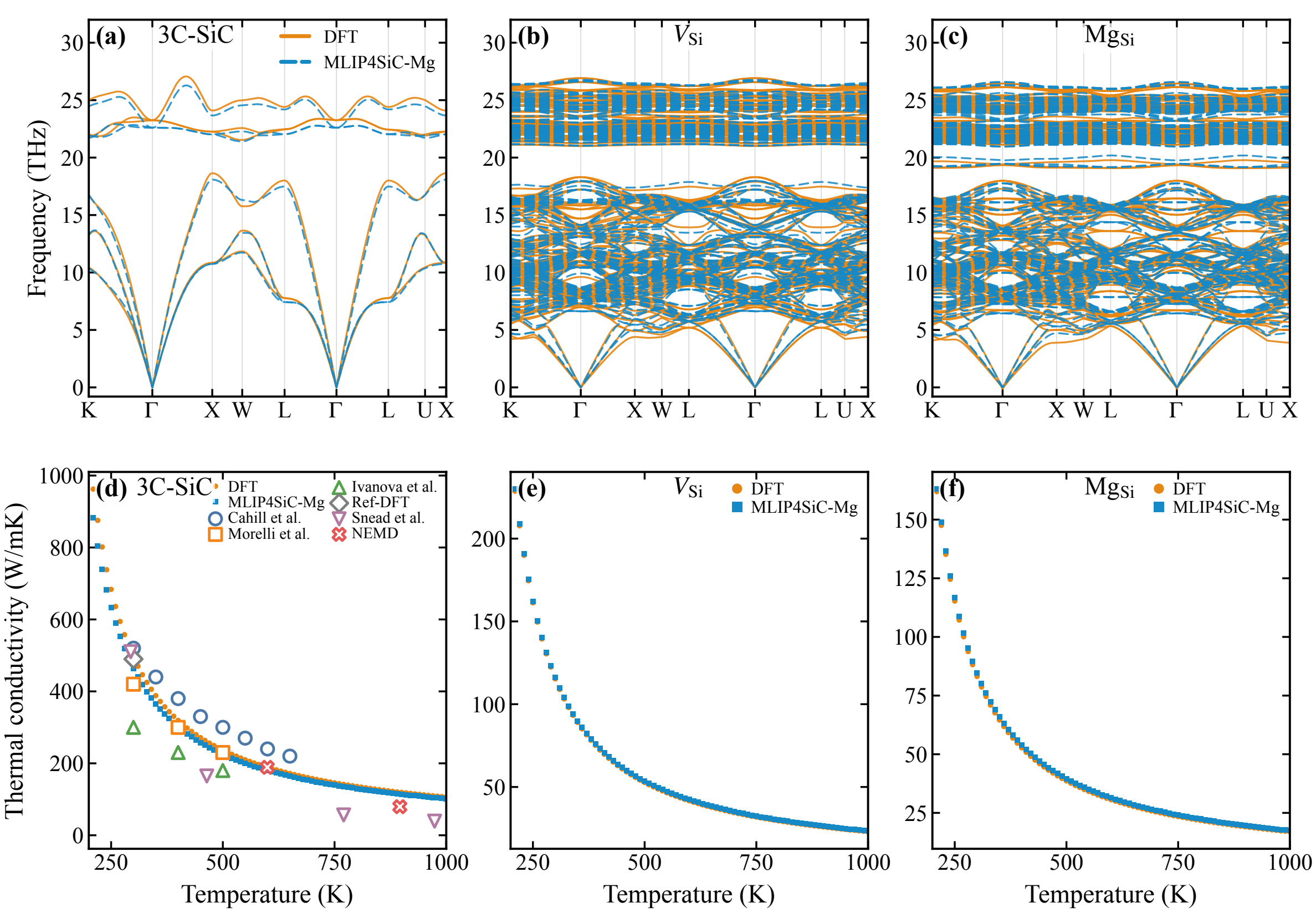

FIG. 5. (a-c) Phonon dispersion curves of 3C-SiC, $V_{Si}$, and $Mg_{Si}$ predicted by MLIP4SiC-Mg and DFT. (d-f) Corresponding lattice thermal conductivities obtained from BTE calculations. Experimental thermal conductivity data for pristine 3C-SiC are taken from Refs. [23–25].

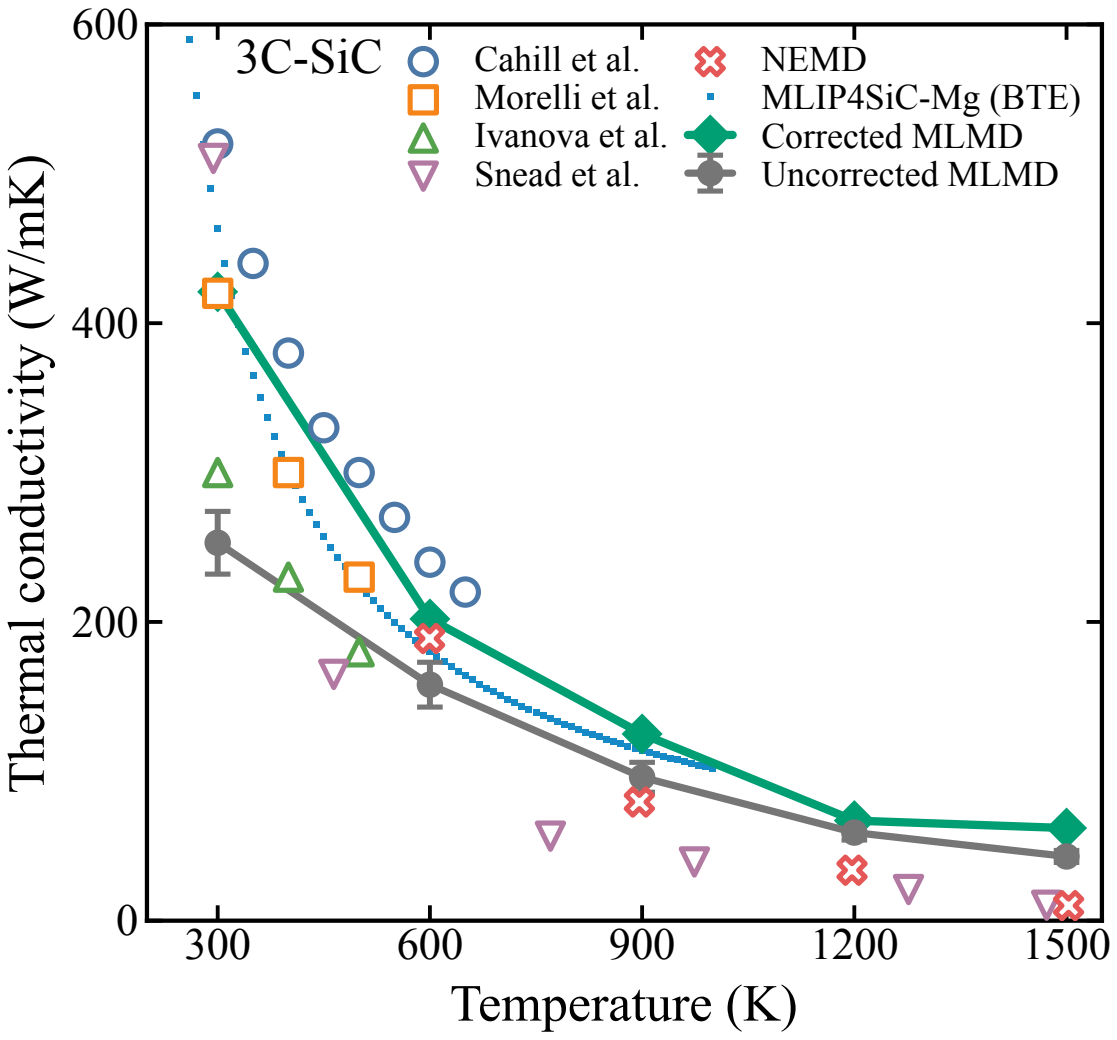

FIG. 6. Corrected thermal conductivity of pristine 3C-SiC obtained using the Langevin-noise extrapolation scheme, compared with the uncorrected MLMD results obtained using the Nosé-Hoover chain thermostat. Experimental data for pristine 3C-SiC are taken from Refs. [23–25].

Among the Mg-containing defects, $Mg_{Si}$ consistently gives the highest thermal conductivity and therefore has the weakest impact on phonon transport. At 0.065 at.%, its thermal conductivity decreases from $\sim$180 $W\,m^{-1}\,K^{-1}$ at 300 K to $\sim$39 $W\,m^{-1}\,K^{-1}$ at 1500 K, and remains above the corresponding values of both $V_C$ and $Mg_{TC}$ over the full temperature range. This behavior indicates that isolated substitutional Mg on the Si sublattice introduces only moderate mass and bonding perturbations. In contrast, $Mg_{TC}$ suppresses thermal conductivity much more strongly. At 0.260 at.%, its thermal conductivity is $\sim$59, 42, 31, 29, and 28 $W\,m^{-1}\,K^{-1}$ at 300, 600, 900, 1200, and 1500 K, respectively, which is comparable to or lower than that of $V_C$ at the same nominal concentration. This result suggests that $Mg_{TC}$ produces substantially stronger local structural distortion and force-constant perturbation than $Mg_{Si}$. The $Mg_{Si}$-$V_C$ complex further lowers the thermal conductivity relative to isolated $Mg_{Si}$, showing that vacancy-associated Mg defects are much more effective phonon scatterers. Overall, the effect of transmutation Mg on heat transport is strongly configuration dependent: isolated substitutional Mg is comparatively benign, whereas non-substitutional or vacancy-associated Mg defects are substantially more detrimental.

For all defect types, increasing defect concentration leads to an additional reduction in thermal conductivity, as expected from the higher density of phonon scattering centers. This concentration dependence is strongest at low temperatures. For example, for $V_C$, increasing the concentration from 0.065 to 0.260 at.% reduces the thermal conductivity from $\sim$128 to $\sim$60 $W\,m^{-1}\,K^{-1}$ at 300 K, corresponding to a decrease of about 53%, whereas at 1500 K the reduction is from $\sim$39 to $\sim$25 $W\,m^{-1}\,K^{-1}$, or about 36%. A similar trend is found for $Mg_{Si}$, for which the conductivity decreases by about 56% at 300 K when the concentration increases from 0.065 to 0.260 at.%, but by only about 26% at 1500 K. This weaker concentration dependence at high temperature is consistent with a Matthiessen-type picture, in which defect scattering adds an approximately temperature-insensitive resistance while intrinsic phonon-phonon scattering increases rapidly with temperature and eventually dominates the total resistance.

It should also be noted that defect concentration is reported here in atomic percent (at.%), defined as the fraction of defective lattice sites. Under this definition, the same nominal at.% does not correspond to the same number density of defect objects for $V_C$ and $Mg_{Si}$-$V_C$. One $V_C$ defect corresponds to one defective site, whereas one $Mg_{Si}$-$V_C$ complex contains two defective sites, namely one substitutional Mg and one C vacancy. At the same at.%, the number density of $V_C$ defects is therefore approximately twice that of $Mg_{Si}$-$V_C$ complexes. The comparison between $V_C$ and $Mg_{Si}$-$V_C$ at equal at.% should therefore be understood as a comparison at equal defective-site fraction, not at equal complex number density. Under this normalization, $V_C$ generally appears more detrimental on a per-defective-site basis. At the same time, the relatively low thermal conductivity of the $Mg_{Si}$-$V_C$ system still highlights the strong phonon-scattering strength of each complex.

### E. Residual thermal resistivity and defect-scattering strength

To further quantify the scattering strength of different defects, we analyze the thermal-conductivity results in terms of defect-induced residual thermal resistivity. For each defect type, the excess thermal resistance relative to pristine 3C-SiC is defined as

$$\Delta R_{\rm def}(T,c) = [\kappa_{\rm def}(T,c)]^{-1} - [\kappa_{\rm SiC}(T)]^{-1}, \qquad (17)$$

where $\kappa_{\rm def}(T,c)$ is the corrected thermal conductivity of the defective system at temperature $T$ and defect concentration $c$, and $\kappa_{\rm SiC}(T)$ is the corrected thermal conductivity of pristine 3C-SiC at the same temperature. An effective residual thermal resistivity per unit defect concentration is then defined as

$$\mathrm{RTR}_{\rm def}(T,c) = \frac{\Delta R_{\rm def}(T,c)}{c}. \qquad (18)$$

Here, $c$ is the defective-site concentration expressed in at.%. If defect scattering followed a strictly dilute Matthiessen-type behavior, $\Delta R_{\rm def}$ would increase linearly with $c$, and $\mathrm{RTR}_{\rm def}$ would be independent of concentration. Deviations from this behavior, therefore, provide a useful measure of nonlinearity in the concentration dependence of defect scattering.

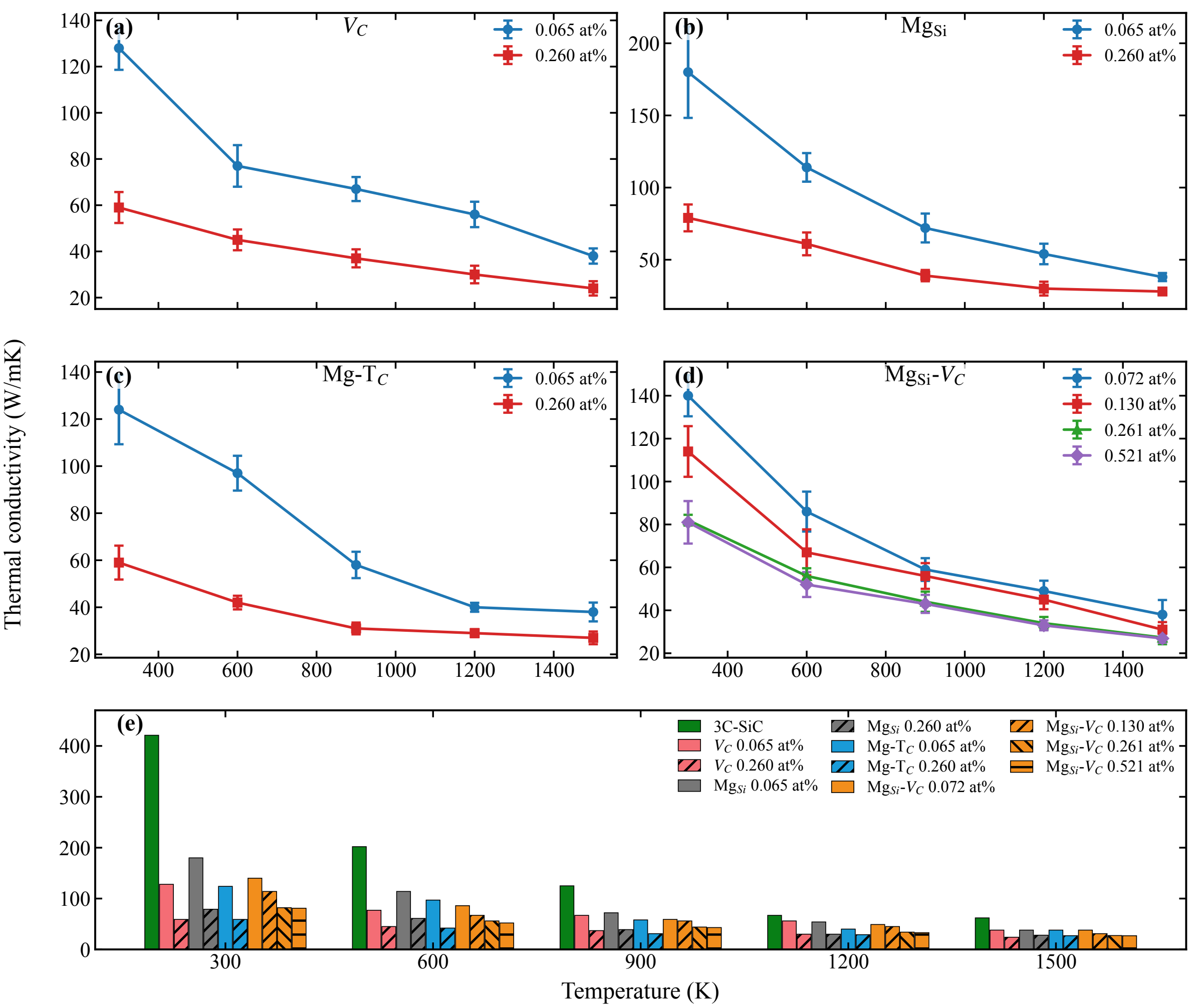

FIG. 7. Corrected thermal conductivities of (a) $V_C$, (b) $Mg_{Si}$, (c) $Mg_{TC}$, and (d) $Mg_{Si}$-$V_C$ at different defect concentrations. (e) Summary of the thermal-conductivity trends for pristine 3C-SiC and the four defective systems as a function of concentration and temperature.

The extracted RTR values, summarized in Tables S4-S8, provide a quantitative measure of the effective phonon-scattering strength of each defect. At 300 K, $V_C$ and $Mg_{TC}$ have the largest RTRs. At 0.065 at.%, their RTRs are $8.38 \times 10^{-2}$ and $8.80 \times 10^{-2}$ m K W$^{-1}$ at.%$^{-1}$, respectively, compared with $4.88 \times 10^{-2}$ m K W$^{-1}$ at.%$^{-1}$ for $Mg_{Si}$. At 0.260 at.%, $V_C$ and $Mg_{TC}$ again give similar RTRs of $5.60 \times 10^{-2}$ and $5.57 \times 10^{-2}$ m K W$^{-1}$ at.%$^{-1}$, both larger than that of $Mg_{Si}$ ($3.98 \times 10^{-2}$ m K W$^{-1}$ at.%$^{-1}$). This ordering supports the physical picture that carbon vacancies and non-substitutional Mg defects introduce stronger phonon scattering than isolated substitutional Mg, because they produce larger local lattice distortion and force-constant perturbations. At higher temperatures, the same general trend remains visible at many concentrations, especially for $Mg_{TC}$, but the differences among defect types become less distinct as intrinsic anharmonic phonon-phonon scattering becomes increasingly important.

The RTR values also show that the defect-induced resistance is not strictly linear in concentration. For $V_C$, increasing the concentration from 0.065 to 0.260 at.% corresponds to a fourfold increase in defective-site concentration, but $\Delta R_{\mathrm{def}}$ increases only from $5.45 \times 10^{-3}$ to $14.56 \times 10^{-3}$ m K W$^{-1}$ at 300 K, from $8.12 \times 10^{-3}$ to $17.07 \times 10^{-3}$ m K W$^{-1}$ at 600 K, and from $7.04 \times 10^{-3}$ to $19.25 \times 10^{-3}$ m K W$^{-1}$ at 900 K. Similar behavior is observed for $Mg_{Si}$ and $Mg_{TC}$ over most of the temperature range. This nonlinearity means that a single concentration-independent residual thermal resistivity cannot fully describe defect scattering over the entire concentration range. Instead, the apparent RTR generally depends on concentration, reflecting changes in the population of heat-carrying phonons that remain available for additional defect scattering. Once the longest-mean-free-path phonons have already been strongly suppressed at lower defect concentration, adding more defects can produce a smaller incremental increase in ther-mal resistance. The deviations are not perfectly monotonic at all temperatures, which is expected because the RTR is extracted from finite-temperature MD data and represents an effective scattering metric rather than a strict dilute-limit constant.

The apparent RTR is also temperature dependent. At the higher defective-site concentration of 0.260 at.%, the RTR of $V_C$ increases from $5.60 \times 10^{-2}$ m K W$^{-1}$ at.%$^{-1}$ at 300 K to $9.77 \times 10^{-2}$ m K W$^{-1}$ at.%$^{-1}$ at 1500 K. The corresponding values increase from $3.98 \times 10^{-2}$ to $7.53 \times 10^{-2}$ m K W$^{-1}$ at.%$^{-1}$ for $Mg_{Si}$, from $5.57 \times 10^{-2}$ to $7.84 \times 10^{-2}$ m K W$^{-1}$ at.%$^{-1}$ for $Mg_{TC}$, and from $3.78 \times 10^{-2}$ to $7.85 \times 10^{-2}$ m K W$^{-1}$ at.%$^{-1}$ for $Mg_{Si}$-$V_C$. This temperature dependence does not imply that the microscopic defect-scattering mechanism is a simple increasing function of temperature. Rather, it reflects the fact that the RTR defined here is an effective resistance-based quantity. As temperature increases, the dominant heat-carrying phonons, their mean free paths, and their relative weighting in the total thermal conductivity all change. In addition, because thermal conductivity becomes smaller at high temperature, differences in inverse conductivity can remain significant even when the absolute thermal conductivities of different defective systems become closer. Therefore, RTR should be interpreted as an effective measure of defect scattering at a given temperature and concentration, rather than as a strictly temperature-independent material constant.

The RTR analysis further clarifies the effect of $Mg_{Si}$-$V_C$ clustering. Because one $Mg_{Si}$-$V_C$ complex contains two defective sites, a defective-site concentration of 0.520 at.% corresponds to an approximate complex-object concentration of 0.260 at.%. This allows a direct comparison with a hypothetical system containing spatially separated $Mg_{Si}$ and $V_C$ defects, each at 0.260 at.%. If the two isolated defects scatter phonons independently, their combined excess thermal resistance can be estimated using a Matthiessen-type sum:

$$\Delta R^{\mathrm{iso}}_{\mathrm{MgSi}+\mathrm{V_C}} = \Delta R_{\mathrm{Mg_{Si}}} + \Delta R_{\mathrm{V_C}}. \tag{19}$$

Using this estimate, the isolated-defect excess resistance is $24.91 \times 10^{-3}$, $28.57 \times 10^{-3}$, $36.86 \times 10^{-3}$, $36.34 \times 10^{-3}$, and $45.00 \times 10^{-3}$ m K W$^{-1}$ at 300, 600, 900, 1200, and 1500 K, respectively. In contrast, the directly calculated $Mg_{Si}$-$V_C$ complex at 0.520 at.% defective-site concentration gives smaller excess resistances of $9.80 \times 10^{-3}$, $14.31 \times 10^{-3}$, $15.50 \times 10^{-3}$, $14.93 \times 10^{-3}$, and $20.96 \times 10^{-3}$ m K W$^{-1}$ over the same temperature range. Thus, forming the $Mg_{Si}$-$V_C$ cluster reduces the total excess thermal resistance relative to two spatially separated defect centers. This reduction likely occurs because the lattice and force-constant perturbations associated with $Mg_{Si}$ and $V_C$ overlap and act as a single composite scattering center, rather than as two independent phonon scatterers.

At the same time, the $Mg_{Si}$-$V_C$ complex should not be regarded as benign. At dilute concentration, its RTR is generally larger than that of isolated $Mg_{Si}$ at low and intermediate temperatures. For example, at approximately 0.07 at.%, the complex gives $\Delta R_{\mathrm{def}} = 4.77 \times 10^{-3}$, $6.64 \times 10^{-3}$, $8.81 \times 10^{-3}$, and $5.36 \times 10^{-3}$ m K W$^{-1}$ at 300, 600, 900, and 1200 K, respectively, all larger than the corresponding $Mg_{Si}$ values of $3.17 \times 10^{-3}$, $3.82 \times 10^{-3}$, $5.85 \times 10^{-3}$, and $3.55 \times 10^{-3}$ m K W$^{-1}$. Therefore, vacancy association enhances the scattering strength of Mg-related defects relative to isolated substitutional Mg, even though the resulting cluster scatters less strongly than the sum of two spatially separated $Mg_{Si}$ and $V_C$ defects.

Taken together, the thermal-resistance and RTR analyses provide a quantitative picture of how Mg transmutation affects thermal transport in 3C-SiC. The effect of Mg cannot be represented by a single generic Mg-defect scattering strength. Instead, it depends on the local defect configuration, defect concentration, and temperature. $V_C$ and $Mg_{TC}$ are generally strong phonon scatterers, while isolated $Mg_{Si}$ is comparatively weak, especially at low and intermediate temperatures. $Mg_{Si}$-$V_C$ clustering increases scattering relative to isolated $Mg_{Si}$, but decreases scattering relative to a Matthiessen sum of spatially separated $Mg_{Si}$ and $V_C$ defects. From the standpoint of thermal-conductivity retention, suppressing vacancy-containing and non-substitutional Mg-related defects is therefore more important than suppressing isolated $Mg_{Si}$ itself.

## IV. CONCLUSION

In this work, we developed a transferable machine-learning interatomic potential, MLIP4SiC-Mg, for 3C-SiC containing intrinsic defects, Mg transmutation products, and Mg-related defect complexes. The model was trained on a large DFT dataset covering pristine SiC structures, SiC polymorphs, intrinsic defects, isolated Mg defects, and Mg-containing complexes. Benchmark calculations show that MLIP4SiC-Mg reproduces DFT energies, forces, equation-of-state behavior, phonon dispersions, and lattice thermal conductivities with near-DFT accuracy. This validation establishes the potential as a reliable tool for thermal-transport simulations of defective SiC systems that are beyond the practical size limits of conventional lattice-dynamics calculations based on the Boltzmann transport equation.

Using Green-Kubo equilibrium molecular dynamics together with force-error correction, we quantified the thermal conductivity of pristine and defective 3C-SiC over a range of temperatures and defect concentrations relevant to irradiation environments. For pristine 3C-SiC, the corrected thermal conductivity at 300 K is 421 W m$^{-1}$ K$^{-1}$, in good agreement with available experimental data. For defective systems, we introduced a resistance-based correction scheme appropriate for the dilute-defect regime. All defect types considered produce a pronounced reduction in thermal conductivity, especially at low temperature, where long-mean-free-path acoustic phonons are highly sensitive to mass disorder, force-constant perturbations, and local lattice distortion. Among the Mg-related defects, isolated substitutional $Mg_{Si}$ has the weakest effect on heat transport, whereas $Mg_{TC}$ and $Mg_{Si}$-$V_C$ are substantially more detrimental.

To quantify defect-scattering strength, we further analyzed the results in terms of defect-induced residual thermal resistivity. This analysis shows that excess thermal resistance is not strictly linear with defect concentration, indicating that the incremental scattering contribution per added defect changes as the population of long-mean-free-path heat carriers is progressively suppressed. The apparent residual thermal resistivity also depends on temperature and should therefore be interpreted as an effective scattering metric at a given temperature and concentration, rather than as a universal constant. On a per-defective-site basis, $V_C$ and $Mg_{TC}$ generally act as strong phonon scatterers, while $Mg_{Si}$ is comparatively weak, particularly at low and intermediate temperatures. The $Mg_{Si}$-$V_C$ complex is more detrimental than isolated $Mg_{Si}$, confirming that vacancy association enhances the scattering strength of Mg-related defects. However, comparison with a Matthiessen-type estimate shows that $Mg_{Si}$-$V_C$ clustering reduces the total excess thermal resistance relative to spatially separated $Mg_{Si}$ and $V_C$ defects, likely because the two local perturbations overlap and act as a single composite scattering center.

Overall, this work clarifies how intrinsic defects, Mg transmutation products, and Mg-defect complexes jointly degrade heat conduction in 3C-SiC. The effect of Mg cannot be treated as a single generic impurity contribution; instead, it depends strongly on the local defect configuration, concentration, and temperature. From the standpoint of thermal-conductivity retention, suppressing vacancy-containing and non-substitutional Mg-related defects is more important than suppressing isolated $Mg_{Si}$ itself. More broadly, the combination of transferable MLIP development, force-error-corrected molecular dynamics, and resistance-based defect analysis provides a general framework for studying thermal transport in irradiation-damaged ceramics.

## DATA AVAILABILITY

The structural training data, potential parameter files, and EMD datasets are openly available through Figshare at `https://doi.org/10.6084/m9.figshare.32667726`. Additional supporting data are available from the corresponding authors upon reasonable request.

## ACKNOWLEDGMENT

This work was funded by the U.S. Department of Energy, Office of Science, Office of Advanced Scientific Computing Research and Office of Fusion Energy Sciences, Scientific Discovery through Advanced Computing (SciDAC) program under Award Number DE-SC0024401. The authors also gratefully acknowledge the computing time provided to them on the high-performance computer Lichtenberg at the NHR Centers NHR4CES at TU Darmstadt. (Project ID: p0026451)

---

[1] Y. Katoh, L. L. Snead, I. Szlufarska, W. J. Weber, Radiation effects in sic for nuclear structural applications, Current Opinion in Solid State and Materials Science 16 (3) (2012) 143–152. doi:10.1016/j.cossms.2012.03.005.

[2] J. Marian, W. Setyawan, Y. Yang, A. Manzoor, W. Zhong, J. R. Trelewicz, J. Yu, E. Peterson, Y. Katoh, L. Snead, et al., Computational materials assessment of the d/li-stripping neutron source as a prototypical facility for fusion materials testing, Current Opinion in Solid State and Materials Science 38 (2025) 101231.

[3] L. L. Snead, T. Nozawa, Y. Katoh, T.-S. Byun, S. Kondo, D. A. Petti, Handbook of sic properties for fuel performance modeling, Journal of nuclear materials 371 (1-3) (2007) 329–377.

[4] P. A. Demkowicz, B. Liu, J. D. Hunn, Coated particle fuel: Historical perspectives and current progress, Journal of Nuclear Materials 515 (2019) 434–450.

[5] K. A. Terrani, L. L. Snead, J. C. Gehin, Microencapsulated fuel technology for commercial light water and advanced reactor application, Journal of Nuclear Materials 427 (1-3) (2012) 209–224.

[6] K. A. Terrani, Accident tolerant fuel cladding development: Promise, status, and challenges, Journal of Nuclear Materials 501 (2018) 13–30.

[7] K. A. Terrani, D. Wang, L. J. Ott, R. O. Montgomery, The effect of fuel thermal conductivity on the behavior of lwr cores during loss-of-coolant accidents, Journal of Nuclear Materials 448 (1-3) (2014) 512–519.

[8] H. L. Heinisch, L. R. Greenwood, W. J. Weber, R. E. Williford, Displacement damage in silicon carbide irradiated in fission reactors, Journal of Nuclear Materials 327 (2–3) (2004) 175–181. doi:10.1016/j.jnucmat.2004.02.012.

[9] T. Koyanagi, Y. Katoh, Mechanical properties of sic composites neutron irradiated under light water reactor relevant temperature and dose conditions, Journal of Nuclear Materials 494 (2017) 46–54. doi:10.1016/j.jnucmat.2017.07.007.

[10] M. E. Sawan, Damage parameters of structural materials in fusion environment compared to fission reactor irradiation, Fusion Engineering and Design 87 (5–6) (2012) 551–555. doi:10.1016/j.fusengdes.2012.01.022.

[11] L. El-Guebaly, M. Sawan, Consequences of neutron energy spectrum on radiation damage, gas production, and transmutations in fusion materials, Fusion Science and Technology 79 (8) (2023) 932–940. doi:10.1080/15361055.2023.2181049.

[12] L. Snead, S. Zinkle, D. White, Thermal conductivity degradation of ceramic materials due to low temperature,

low dose neutron irradiation, Journal of nuclear materials 340 (2-3) (2005) 187–202.

[13] M. Rohde, Reduction of the thermal conductivity of sic by radiation damage, Journal of nuclear materials 182 (1991) 87–92.

[14] L. Snead, Limits on irradiation-induced thermal conductivity and electrical resistivity in silicon carbide materials, Journal of Nuclear Materials 329 (2004) 524–529.

[15] M. A. Pickering, R. L. Taylor, J. T. Keeley, G. A. Graves, Chemically vapor deposited silicon carbide (sic) for optical applications, Nuclear Instruments and Methods in Physics Research Section A: Accelerators, Spectrometers, Detectors and Associated Equipment 291 (1-2) (1990) 95–100.

[16] R. H. Jones, L. Giancarli, A. Hasegawa, Y. Katoh, A. Kohyama, B. Riccardi, L. L. Snead, W. J. Weber, Promise and challenges of sicf/sic composites for fusion energy applications, Journal of Nuclear Materials 307 (2002) 1057–1072.

[17] M. Sawan, Y. Katoh, L. L. Snead, Transmutation of silicon carbide in fusion nuclear environment, Journal of Nuclear Materials 442 (1-3) (2013) S370–S375.

[18] W. Jiang, H. J. Jung, L. Kovarik, Z. Wang, T. J. Roosendaal, Z. Zhu, D. J. Edwards, S. Hu, C. H. J. Henager, R. J. Kurtz, Y. Wang, Magnesium behavior and structural defects in $mg^+$ ion implanted silicon carbide, Journal of Nuclear Materials 458 (2015) 146–155. doi:10.1016/j.jnucmat.2014.12.071.

[19] K.-Y. Lim, Y.-W. Kim, K. J. Kim, Mechanical properties of electrically conductive silicon carbide ceramics, Ceramics International 40 (7) (2014) 10577–10582.

[20] S. Kultayeva, J.-H. Ha, R. Malik, Y.-W. Kim, K. J. Kim, Effects of porosity on electrical and thermal conductivities of porous sic ceramics, Journal of the European Ceramic Society 40 (4) (2020) 996–1004.

[21] H. W. Li, Y. P. Zhao, G. Q. Chen, M. H. Li, Z. F. Wei, X. S. Fu, W. L. Zhou, Sic-based ceramics with remarkable electrical conductivity prepared by ultrafast high-temperature sintering, Journal of the European Ceramic Society 43 (5) (2023) 2269–2274.

[22] E. Hodgson, M. Malo, J. Manzano, A. Moroño, T. Hernandez, Radiation induced modification of electrical conductivity for three types of sic, Journal of nuclear materials 417 (1-3) (2011) 421–424.

[23] Z. Cheng, J. Liang, K. Kawamura, H. Zhou, H. Asamura, H. Uratani, J. Tiwari, S. Graham, Y. Ohno, Y. Nagai, et al., High thermal conductivity in wafer-scale cubic silicon carbide crystals, Nature communications 13 (1) (2022) 7201.

[24] D. Morelli, J. Heremans, C. Beetz, W. Woo, G. Harris, C. Taylor, Carrier concentration dependence of the thermal conductivity of silicon carbide, in: Institute of Physics Conference Series, Vol. 137, Bristol [England]; Boston: Adam Hilger, Ltd., c1985-, 1994, pp. 313–316.

[25] L. Ivanova, P. Aleksandrov, K. Demakov, Thermoelectric properties of vapor-grown polycrystalline cubic sic, Inorganic materials 42 (11) (2006) 1205–1209.

[26] C. Shen, S. Attarian, Y. Zhang, H. Zhang, M. Asta, I. Szlufarska, D. Morgan, Supersalt: equivariant neural network force fields for multicomponent molten salts system, Nature Communications 16 (1) (2025) 7280.

[27] R. Jacobs, D. Morgan, S. Attarian, J. Meng, C. Shen, Z. Wu, C. Y. Xie, J. H. Yang, N. Artrith, B. Blaiszik, et al., A practical guide to machine learning interatomic potentials–status and future, Current Opinion in Solid State and Materials Science 35 (2025) 101214.

[28] B. Fu, Y. Sun, W. Jiang, F. Wang, L. Zhang, H. Wang, B. Xu, Determining the thermal conductivity and phonon behavior of sic materials with quantum accuracy via deep learning interatomic potential model, Journal of Nuclear Materials 591 (2024) 154897.

[29] H. Zhang, M. Cheng, X. Jiang, H. Zhang, X. Pi, D. Yang, T. Deng, Neuroevolution potential for thermal transport in silicon carbide, Journal of Materials Informatics 5 (3) (2025) N–A.

[30] W. Liu, P. Guo, Z. Zheng, S. Chen, Y.-N. Wu, Neural-network potential for defect formation induced by knock-on irradiation damage in 4h-sic, Advanced Electronic Materials (2025) 2400911.

[31] X. Wu, W. Zhou, H. Dong, P. Ying, Y. Wang, B. Song, Z. Fan, S. Xiong, Correcting force error-induced underestimation of lattice thermal conductivity in machine learning molecular dynamics, The Journal of Chemical Physics 161 (1) (2024).

[32] J. P. Perdew, K. Burke, M. Ernzerhof, Generalized gradient approximation made simple, Physical review letters 77 (18) (1996) 3865.

[33] G. Kresse, J. Furthmüller, Efficient iterative schemes for ab initio total-energy calculations using a plane-wave basis set, Physical review B 54 (16) (1996) 11169.

[34] P. E. Blöchl, Projector augmented-wave method, Physical review B 50 (24) (1994) 17953.

[35] Z. Fan, Z. Zeng, C. Zhang, Y. Wang, K. Song, H. Dong, Y. Chen, T. Ala-Nissila, Neuroevolution machine learning potentials: Combining high accuracy and low cost in atomistic simulations and application to heat transport, Physical Review B 104 (10) (2021) 104309.

[36] H. Dong, Y. Shi, P. Ying, K. Xu, T. Liang, Y. Wang, Z. Zeng, X. Wu, W. Zhou, S. Xiong, et al., Molecular dynamics simulations of heat transport using machine-learned potentials: A mini-review and tutorial on gpumd with neuroevolution potentials, Journal of Applied Physics 135 (16) (2024).

[37] S. Plimpton, Fast parallel algorithms for short-range molecular dynamics, Journal of computational physics 117 (1) (1995) 1–19.

[38] T. Liang, K. Xu, E. Lindgren, Z. Chen, R. Zhao, J. Liu, E. Berger, B. Tang, B. Zhang, Y. Wang, et al., Nep89: Universal neuroevolution potential for inorganic and organic materials across 89 elements, arXiv preprint arXiv:2504.21286 (2025).

[39] P. K. Schelling, S. R. Phillpot, P. Keblinski, Comparison of atomic-level simulation methods for computing thermal conductivity, Physical Review B 65 (14) (2002) 144306.

[40] J. Li, L. Porter, S. Yip, Atomistic modeling of finite-temperature properties of crystalline $\beta$-SiC: Ii. thermal conductivity and effects of point defects, Journal of Nuclear Materials 255 (2-3) (1998) 139–152. doi:10.1016/S0022-3115(98)00034-8.

[41] T. Kawamura, D. Hori, Y. Kangawa, K. Kakimoto, M. Yoshimura, Y. Mori, Thermal conductivity of SiC calculated by molecular dynamics, Japanese Journal of Applied Physics 47 (12) (2008) 8898–8901. doi:10.1143/JJAP.47.8898.

[42] J.-P. Crocombette, G. Dumazer, N. Q. Hoang, F. Gao, W. J. Weber, Molecular dynamics modeling of the thermal conductivity of irradiated SiC as a function of cas-

cade overlap, Journal of Applied Physics 101 (2) (2007) 023527. doi:10.1063/1.2431397.

[43] G. D. Samolyuk, S. I. Golubov, Y. N. Osetsky, R. E. Stoller, Molecular dynamics study of influence of vacancy types defects on thermal conductivity of $\beta$-SiC, Journal of Nuclear Materials 418 (1-3) (2011) 174–181. doi:10.1016/j.jnucmat.2011.06.036.

[44] Q. Wang, N. Gui, X. Huang, X. Yang, J. Tu, S. Jiang, The effect of temperature and cascade collision on thermal conductivity of 3C-SiC: A molecular dynamics study, International Journal of Heat and Mass Transfer 180 (2021) 121822. doi:10.1016/j.ijheatmasstransfer.2021.121822.

[45] K. Yin, L. Shi, X. Ma, Y. Zhong, M. Li, X. He, Thermal conductivity of 3C/4H-SiC nanowires by molecular dynamics simulation, Nanomaterials 13 (15) (2023) 2196. doi:10.3390/nano13152196.

[46] N. H. Protik, A. Katre, L. Lindsay, J. Carrete, N. Mingo, D. Broido, Phonon thermal transport in 2H, 4H and 6H silicon carbide from first principles, Materials Today Physics 1 (2017) 31–38. doi:10.1016/j.mtphys.2017.05.004.

[47] A. Katre, J. Carrete, B. Dongre, G. K. H. Madsen, N. Mingo, Exceptionally strong phonon scattering by B substitution in cubic SiC, Physical Review Letters 119 (7) (2017) 075902. doi:10.1103/PhysRevLett.119.075902.

[48] Z. Cheng, J. Liang, K. Kawamura, H. Zhou, H. Asamura, H. Uratani, J. Tiwari, S. Graham, Y. Ohno, Y. Nagai, T. Feng, N. Shigekawa, D. G. Cahill, High thermal conductivity in wafer-scale cubic silicon carbide crystals, Nature Communications 13 (2022) 7201. doi:10.1038/s41467-022-34943-w.

[49] J. E. Turney, A. J. H. McGaughey, C. H. Amon, Assessing the applicability of quantum corrections to classical thermal conductivity predictions, Physical Review B 79 (22) (2009) 224305. doi:10.1103/PhysRevB.79.224305.

[50] O. N. Bedoya-Martínez, J.-L. Barrat, D. Rodney, Computation of the thermal conductivity using methods based on classical and quantum molecular dynamics, Physical Review B 89 (1) (2014) 014303. doi:10.1103/PhysRevB.89.014303.

[51] M. Puligheddu, Y. Xia, M. Chan, G. Galli, Computational prediction of lattice thermal conductivity: A comparison of molecular dynamics and boltzmann transport approaches, Physical Review Materials 3 (8) (2019) 085401. doi:10.1103/PhysRevMaterials.3.085401.

[52] H. Zhou, S. Zhou, Z. Hua, K. Bawane, T. Feng, Impact of classical statistics on thermal conductivity predictions of BAs and diamond using machine learning molecular dynamics, Applied Physics Letters 125 (17) (2024) 172202. doi:10.1063/5.0238592.

[53] Y. Liu, H. Wang, L. Guo, Z. Yan, J. Zheng, W. Zhou, J. Xue, Deep learning inter-atomic potential for irradiation damage in 3c-sic, Computational Materials Science 233 (2024) 112693.

[54] A. Jain, J. Montoya, S. Dwaraknath, N. E. Zimmermann, J. Dagdelen, M. Horton, P. Huck, D. Winston, S. Cholia, S. P. Ong, et al., The materials project: Accelerating materials design through theory-driven data and tools, in: Handbook of Materials Modeling: Methods: Theory and Modeling, Springer, 2020, pp. 1751–1784.

[55] L. Zhang, W. Luo, R. Liu, M. Chen, Z. Yan, K. Cao, Exploring the energy landscape of aluminas through machine learning interatomic potential, Physical Review Materials 9 (2) (2025) 023801.

[56] T. Liang, P. Ying, K. Xu, Z. Ye, C. Ling, Z. Fan, J. Xu, Mechanisms of temperature-dependent thermal transport in amorphous silica from machine-learning molecular dynamics, Physical Review B 108 (18) (2023) 184203.

[57] J. Zhao, J. Byggmästar, H. He, K. Nordlund, F. Djurabekova, M. Hua, Complex ga2o3 polymorphs explored by accurate and general-purpose machine-learning interatomic potentials, NPJ Computational Materials 9 (1) (2023) 159.

[58] K. Karch, F. Bechstedt, P. Pavone, D. Strauch, Pressure-dependent properties of sic polytypes, Physical Review B 53 (20) (1996) 13400.

[59] M. Yoshida, A. Onodera, M. Ueno, K. Takemura, O. Shimomura, Pressure-induced phase transition in sic, Physical Review B 48 (14) (1993) 10587.

[60] W. Bassett, M. Weathers, T.-C. Wu, T. Holmquist, Compressibility of sic up to 68.4 gpa, Journal of applied physics 74 (6) (1993) 3824–3826.

[61] H. Liu, A. Reza, F. Hofmann, S. de Moraes Shubeita, G. Glodan, R. Harrison, Thermal conductivity degradation of silicon-ion-irradiated silicon carbide ceramics, Acta Materialia (2025) 121329.

[62] R. Ullah, D. Morgan, I. Szlufarska, Mg and native defects in cubic silicon carbide from first principles, Journal of Physics D: Applied Physics (2025).

[63] T. Fan, W. Liu, Z. Ruan, Y. Cao, T. Ye, J. Liu, F. Zhong, X. Tan, H. Liang, D. Chen, et al., First-principles investigation of effects of defects on the physical properties of 3c-sic under high temperatures and pressures, Journal of Materials Research and Technology 20 (2022) 3633–3645.

[64] P. Korotaev, I. Novoselov, A. Yanilkin, A. Shapeev, Accessing thermal conductivity of complex compounds by machine learning interatomic potentials, Physical Review B 100 (14) (2019) 144308.

[65] X. Qian, S. Peng, X. Li, Y. Wei, R. Yang, Thermal conductivity modeling using machine learning potentials: application to crystalline and amorphous silicon, Materials Today Physics 10 (2019) 100140.

[66] C. Shen, M. Dai, X. Xiao, N. Hadaeghi, W. Xie, A. Weidenkaff, T. Tadano, H. Zhang, Impact of quartic anharmonicity on lattice thermal transport in eutio3: A comparative theoretical and experimental investigation, Materials Today Physics 34 (2023) 101059.

[67] R. Yan, C. Shen, M. Widenmeyer, T. Luo, R. Winkler, E. Adabifiroozjaei, R. Xie, S. Yoon, E. Suard, L. Molina-Luna, et al., The role of interstitial cu on thermoelectric properties of zrnisn half-heusler compounds, Materials Today Physics 33 (2023) 101049.